*Review*

# Review of Novel Approaches to Organic Liquid Scintillators in Neutrino Physics

Stefan Schoppmann

Detektorlabor, Exzellenzcluster PRISMA+, Johannes Gutenberg-Universität Mainz, Staudingerweg 9, 55128 Mainz, Germany; stefan.schoppmann@uni-mainz.de

**Abstract:** Organic liquid scintillators have been used for decades in many neutrino physics experiments. They are particularly suited for the detection of low-energy neutrinos where energy and timing information is required. Organic liquid scintillators exhibit advantages such as high light yield, cost effectiveness, radio purity, and more. However, they also entail disadvantages, most prominently a lack of vertex resolution and particle identification. In recent years, various novel ideas have emerged to improve the performance of organic liquid scintillators. In this review, novel approaches to organic liquid scintillators in neutrino experiments as of 2022 are reviewed and their prospects and applications compared.

**Keywords:** liquid scintillators; neutrino physics; light detection; detector design

## 1. Introduction

Organic liquid scintillators have been a key technology in the field of neutrino physics for decades. They are especially suited for low-energy neutrino applications due to their high light output and proportional response to the incident particle energy. The first experiment to successfully detect neutrinos already used a liquid scintillator in 1956 [1]. Since then, they have been used in numerous detectors due to their high purity, low energy threshold, volume flexibility and scalability, low costs, and homogeneity.

In the field of neutrino physics and related fields of research, organic liquid scintillators have allowed for several measurements and discoveries. These include the understanding of neutrino flavor mixing and oscillations through long baseline observations by KamLAND [2] and NOvA [3], as well as short baseline observations from Daya Bay [4], Double Chooz [5], and RENO [6]. KamLAND [7] and later Borexino [8] were able to detect geoneutrinos from Earth. Solar neutrinos detected in Borexino allowed insights into our sun [9,10]. Organic liquid scintillator detectors such as the accelerator-based LSND [11] and the very-short-baseline reactor-based experiments NEOS [12], STEREO [13], PROSPECT [14], and Neutrino-4 [15], investigated indications for additional sterile neutrino states, which are currently under dispute [16,17]. Furthermore, at a very short baseline, nuclear reactor monitoring was achieved by the Nucifer detector [18] using an organic liquid scintillator.

Upcoming organic liquid scintillator detectors, including JUNO [19], Theia [20], KAMland-ZEN [21], SNO+ [22], and Prospect-II [23], can give further insight into our Sun and Earth, supernovae, the Majorana character of neutrinos, neutrino masses, and the existence of additional sterile neutrinos, and allow for improved reactor monitoring [24–26].

To allow for such discoveries, various ideas on the advancement of organic liquid scintillators have been developed in recent years. They mostly target the improvement of individual aspects of organic liquid scintillators by the introduction of new materials into the scintillator or combining the scintillator with other materials. These aspects include improvements to the directional resolution, vertex resolution, particle identification, light yield, metal loading, safety, and radiopurity. While recent developments show promising progress in terms of directional resolution and particle identification with traditional organic liquid scintillators [27,28], the new ideas seek to outperform these achievements. Based on several of those ideas, experimental collaborations have been formed to demonstrate and advance suitable detectors and investigate the performance and discovery potential of these new technologies.

In this article, the basic principles of all organic liquid scintillators will be reviewed in Section 2. In the following sections, modifications and new concepts in organic liquid scintillator technologies will be discussed. After established methods such as metal loading (Section 3) and the blending of scintillators (Section 4) are presented, novel approaches such as low temperature scintillators (Section 5), water-based scintillators (Section 6), scintillators with slow light output (Section 7), opaque scintillators (Section 8), siloxane-based scintillators (Section 9), and quantum dots (Section 10) will be reviewed. For each novel approach, collaborative efforts around this technology will be discussed within its respective section, should such an effort exist. After a new detector concept of floating scintillators is discussed in Section 11, a summary will be given in Section 12.

## 2. Principles of Organic Liquid Scintillators

The main objective of a scintillator is the conversion of the kinetic energy of an incident particle into detectable light. The light emission can therefore be understood as a form of luminescence. For organic liquid scintillators, the key component in this process is the phenyl groups within the scintillator molecules. Several processes within the scintillator on the microphysical scale, as well as incident particle properties, determine the specifics of the scintillation light output, including the wavelength spectrum, the time profile, and the conversion efficiency. In the following subsections, these individual aspects will be discussed in greater detail for organic liquid scintillators.

### 2.1. Mechanism of Scintillation

The primary cause of the scintillation light production is the ionization of scintillator molecules by the incident particle. This is caused either by the particle itself, if it carries electrical charge (e.g., an electron, muon or proton), or indirectly, after the incident particle transfers energy via an interaction to an electrically charged particle present within the scintillator (e.g., neutrino, neutron, and gamma). In both cases, the charged particle(s) are (is) able to ionize or excite the molecules of the scintillator along its path, thereby loosing some of its kinetic energy.

### 2.2. Scintillator Energy Levels

In organic scintillators, the energy transferred to conjugated phenyl groups is of special interest. The phenyl groups contain delocalized electrons in their $\pi$-bonds, which arise from hybridization of the atomic s- and p-orbitals of the carbon atoms that make up the group [29]. Solving Schrödinger's equation for these circular $\pi$-orbitals under the condition of a cyclic wavefunction $\psi(x) = \psi(x + l)$, where $x$ denotes the location and $l$ the circumference, leads to energy states in the 3 to 6 eV range, i.e., blue to ultraviolet photons [30]. Thus, phenyl groups allow an emission of light, so-called fluorescence light, whose wavelength spectrum is suitable for efficient detection with optical sensors. The exact absorption and emission spectra are dependent on the full molecule structure.

### 2.3. Excitation and De-Excitation Processes

In a scintillator molecule, each electronic energy state comprises several vibrational sub-states of much smaller energy splitting than the electronic states. Usually, the excitation of a $\pi$-electron happens from the vibrational ground state of the electronic ground state into an excited vibrational state of an excited electronic state. This is know as the Franck–Condon principle [31,32], stating that during an electronic transition, a change from one vibrational energy level to another is more likely, if the two vibrational wavefunctions have a larger overlap. This is owed to the fact that electronic transitions are practically instantaneous compared to the time scale of nuclear motions. Thus, the new vibrational level must be instantaneously compatible with the nuclear positions and momenta of the vibrational state of the previous electronic state. The de-excitation of vibrational states happens without radiation at a much faster time scale of $10^{-12}$ to $10^{-11}$ s than the subsequent de-excitation of the electronic state. Here, the time scale depends on whether the excited electronic state belongs to the singlet (spin quantum number equals zero) or triplet (spin quantum number equals the unit) regime. For singlet states, times of a few to tens of nanoseconds are typical (fluorescence). For triplet states, times of milliseconds or longer are observed (phosphorescence) because the triplet annihilation reaction involves two excited molecules [29,33].

The decay of excited electronic states in a scintillator component follows a double-exponential relation. There are two time constants $\tau_f$ and $\tau_s$, one describing the fast and the other describing the slow component, participating in the decay:

$$N(t) = A_f \exp\left(-\frac{t}{\tau_f}\right) + A_s \exp\left(-\frac{t}{\tau_s}\right) \tag{1}$$

Here, $t$ denotes time, $N$ is the number of emitted photons, and $A_f$ and $A_s$ are the weight factors of the fast and slow components, respectively. Typical values for the fast and slow time constant in a liquid scintillator are a few and tens of nanoseconds, respectively [34–37]. The weight between fast and slow component is typically at a ratio of 70:30 [29,33]. The time constants and weight factors determine the pulse shape of the events. A precise technique to measure the pulse shape of a given scintillator is the sampling of single scintillation photons in coincidence with a scintillator excitation [38]. In a realistic scintillator with one or more fluors, more than two time constants exist. The fastest time constant in a realistic scintillator is mostly linked to the primary fluor. Because it also includes the time for energy transfer from the solvent to the fluor, it can be reduced to some extent by increasing the fluor concentration until it approaches the intrinsic time constant of the fluor itself. The longer components are mostly attributed to de-excitation of triplet states [39].

*2.4. Fluors and Wavelength Shifters*

In a scintillator with just one solvent, most of the fluorescent radiation is self-absorbed due to a significant overlap of absorption and emission spectra. The quantum yield of such a solvent is typically less than 50%. It is noteworthy that self-absorption does not occur for inorganic liquid scintillators made from noble gases, where scintillation light is produced via the formation of excited dimers. Their excitation levels differ from the levels of the single atoms in the liquid noble gas. To prevent losses due to self-absorption in an organic scintillator, one or more types of fluorescence molecules are introduced into the solvent. They are usually denoted as (primary) fluors or (secondary) wavelength shifters. Primary fluors have typical concentrations of $\mathcal{O}(10^{-3}\text{g/g})$, while secondary wavelength shifters are used at about $\mathcal{O}(10^{-6}\text{g/g})$. The absorption spectrum of the primary fluor shows a significant overlap with the emission spectrum of the solvent. Ideally, the shift of the emission spectrum to longer wavelengths (Stokes shift) allows a more transparent region of the scintillator to be reached.

The energy transfer from the solvent to the fluor is mainly radiationless and happens via Förster resonant energy transfer (FRET) or Dexter electron exchange (DEE) [40,41]. In these processes, either energy is transferred via virtual photons through dipole–dipole coupling between solvent and fluor or an excited electron is transferred directly from the solvent molecule to the fluor. These effects are strongly dependent on the distances between participating molecules and therefore require certain concentrations of fluors in the scintillator. In contrast, too high concentrations of fluors can again lead to self-absorption of the fluor, especially in large volumes. This can be counteracted by introducing more than one fluor, where the absorption spectrum of the second fluor matches the emission of the first fluor, i.e., by daisy-chaining the fluors (cf. Figure 1). Energy transfer after the primary fluor is usually radiative. Thus, a high fluorescence quantum yield of the fluor is decisive, as it determines its re-emission probability. Typical values of common fluors are above 80% and may be dependent on concentration and solvent type [42]. In detector simulations, the use of an effective emission spectrum after the non-radiative transfer is typically simpler, as it avoids simulating complicated microphysics [43].

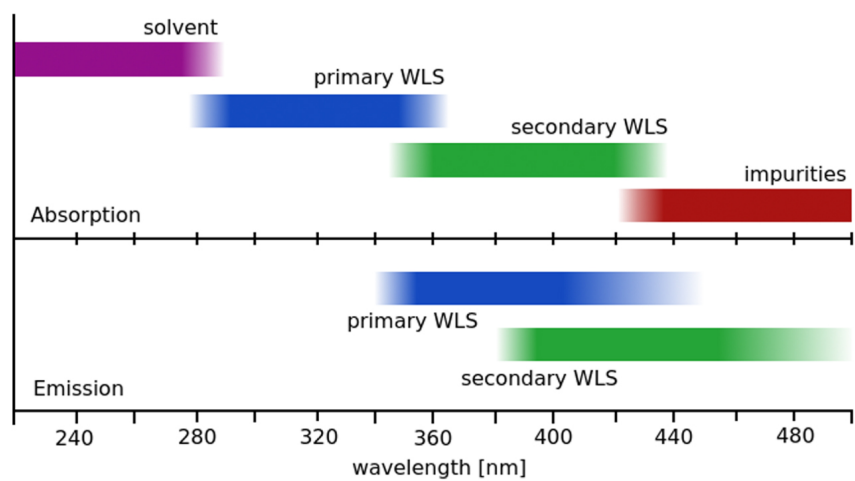

**Figure 1.** Typical arrangement of absorption and emission regions in a solvent with two wavelength shifters (WLS). Reprinted from [39], © 2016.

Due to the high concentration of the solvent compared to that of the fluor, there are typically several energy transfers between solvent molecules before energy is transferred to a fluor molecule ("hopping") [44]. Furthermore, excited molecules can move by diffusion, which causes the scintillation process to be dependent on the viscosity and temperature of the solvent. The increased rate of collisions of solvent molecules at higher temperatures causes a scintillator to have a reduced light output.

*2.5. Light Yield, Quenching, and Pulse Shape Discrimination*

In general, quenching describes any form of loss of deposited energy that decreases the fluorescence intensity of a scintillator [29]. A simple form of quenching is "impurity quenching", where some of the energy of the excited molecules is transferred to non-fluorescent impurities. This energy is then lost for scintillation. However, if not noted otherwise, quenching in the context of liquid scintillators usually refers to the process of "ionization quenching".

In an ideal case, excited and ionized scintillator molecules along the path of an incident particle are sufficiently far apart, such that interactions between them can be neglected. In this ideal case, the energy $L$ emitted as scintillation light is proportional to the particle energy $E$. Typically, 3 to 12% of the incident energy is converted to scintillation light. This leads to a typical light yield of $\mathcal{O}(10^4$ photons/MeV).

In a more realistic case, particle- and energy-dependent losses of scintillation light occur because the density of ionized and excited molecules is rather high. This leads to a non-linear behavior in the scintillation response. The quenching effect is stronger if the energy loss per path length $dE/dx$ is large. This behavior occurs especially at low energies, e.g., towards the end of a particle track. Furthermore, heavy particles exhibit a stronger quenching effect. The light yield of an alpha particle can be more than a factor 10 lower compared to an electron of the same energy. In addition to the higher quenching, more triplet states are also excited, impacting the decay time of the scintillator. From this, the possibility of particle identification via pulse shape discrimination (PSD) arises [33]. While one could take into account the entire time spectrum of the scintillation pulse, it is usually sufficient to compare the late light ratio, i.e., the fraction between the integrated late part of the light pulse and the total light pulse of the scintillation [33]. When following this strategy, the discrimination power improves for high-energy particles, as the distributions of the late light ratio narrow. In addition, removing dissolved oxygen from the scintillator improves its discrimination power [35].

A popular empirical equation to describe the light yield per path length as a function of the energy loss per path length as caused by ionization quenching is Birks' law [29]:

$$\frac{dL}{dx} = S \frac{\frac{dE}{dx}}{1 + k_B \frac{dE}{dx}} \qquad (2)$$

where $dL$ denotes the light produced by a particle along a path of length $dx$, $S$ is the light yield constant, and $k_B$ is Birks' quenching parameter. For electrons, typical values of $k_B$ are about 0.01–0.03 cm/MeV [45]. While this equation is sufficient for most applications, Chou's Equation [46], a more generalized form using two quenching parameters $A$ and $B$, can be used for a more precise description:

$$\frac{dL}{dx} = S \frac{\frac{dE}{dx}}{1 + A\frac{dE}{dx} + B\left(\frac{dE}{dx}\right)^2} \qquad (3)$$

There are other models that describe the quenching mechanisms on a more complex level [47,48]. The models mainly disagree for highly charged particles or high energies [45].

*2.6. Liquid Scintillator Materials*

Liquid scintillators typically comprise one or more solvents with the addition of one, and in some applications a secondary, fluor [39]. Popular choices are alkyl benzenes such as xylene, toluene, or cumene [49]. In particular, pseudocumene (PC, 1,2,4-trimethylbenzene) is know for its high light yield [50–55]. However, material compatibility, e.g., with acrylic vessels, can make PC or other chemically aggressive solvents a poor choice of material. Where safety considerations are significant (e.g., nuclear reactor sites), solvents with high flash points such as linear alkyl benzene (LAB), phenyl xylylethane (PXE), or di-isopropyl naphthalene (DIN) are a common option [36,56–66]. Solvents can also be combined with additions of anti-static agents to prevent sparking and ignition of scintillators during pumping. Among the safe solvents, LAB received special attention since its first consideration in the SNO+ experiment [67]. It excels in its safety features, transparency, compatibility, and costs. LAB is used as basis in many applications discussed in this review.

A common primary fluor is 2,5-diphenyloxazole (PPO) [18,51,52,55–57,62–66,68]. Other fluors with similar optical properties to PPO include 2-(4-biphenyl)-5-phenyl-1,3,4-oxadiazole (PBD), 2-(4-biphenyl)-5-(4-tert-butyl-phenyl)-1,3,4-oxadiazole (butyl-PBD), 2-(4-biphenyl)-5-phenyloxazole (BPO), and p-terphenyl (p-TP) [42,69–72]. The solubility of p-TP is limited in several solvents. BPO has limited transparency in the relevant region and is toxic.

As secondary wavelength shifters, 4-bis-(2-methyl-styryl)-benzene (bis-MSB), 1,4-bis(5-phenyloxazol-2-yl)-benzene (POPOP), and 1-phenyl-3-mesityl-2-pyrazoline (PMP) are often used [18,42,51,56,57,62–66,68,72]. bis-MSB performs well with PPO and p-TP and has a shorter decay time than POPOP. PMP is known for its large Stokes shift.

If the Stokes shift of the primary fluor is large enough, the use of a secondary wavelength shifter can, in principle, be omitted [73]. Table 1 gives an overview of the scintillator components mentioned above.

**Table 1.** Molecular formular, density, and flash point, as well as the wavelengths of the optical absorption and emission peaks for several solvents and fluors diluted in cyclohexane. Reprinted with minor edits from [39], © 2016.

| Molecule | Formula | Density | Flash Point | abs. max. | em. max. |
|---|---|---|---|---|---|
| PC | $C_9H_{12}$ | 0.88 kg/L | 48 °C | 267 nm | 290 nm |
| toluene | $C_7H_8$ | 0.87 kg/L | 4 °C | 262 nm | 290 nm |
| anisole | $C_7H_8O$ | 0.99 kg/L | 43 °C | 271 nm | 293 nm |
| LAB | $C_{6+n}H_{6+2n}$ | 0.87 kg/L | ~140 °C | 260 nm | 284 nm |
| DIN | $C_{16}H_{20}$ | 0.96 kg/L | >140 °C | 279 nm | 338 nm |
| o-PXE | $C_{16}H_{18}$ | 0.99 kg/L | 167 °C | 269 nm | 290 nm |
| n-dodecane | $C_{12}H_{26}$ | 0.75 kg/L | 71 °C | – | – |
| mineral oil | – | ~0.85 kg/L | >130 °C | – | – |
| PPO | $C_{15}H_{11}NO$ | – | – | 303 nm | 358 nm |
| PBD | $C_{20}H_{14}N_2O$ | – | – | 302 nm | 358 nm |
| butyl-PBD | $C_{24}H_{14}N_2O$ | – | – | 302 nm | 361 nm |
| BPO | $C_{21}H_{15}NO$ | – | – | 320 nm | 384 nm |
| p-TP | $C_{18}H_{14}$ | – | – | 276 nm | 338 nm |
| bis-MSB | $C_{24}H_{22}$ | – | – | 345 nm | 418 nm |
| TBP | $C_{28}H_{22}$ | – | – | 347 nm | 455 nm |
| POPOP | $C_{24}H_{16}N_2O_2$ | – | – | 360 nm | 411 nm |
| PMP | $C_{18}H_{20}N_2$ | – | – | 295 nm | 425 nm |

*2.7. Optical Properties*

The attenuation of light in a liquid scintillator can take place via two processes. Light can be absorbed or scattered. The intensity $I$ of a one-dimensional monochromatic light beam decreases exponentially with the attenuation length $\Lambda$. This leads to the definition of absorbance $A$ after a beam has traveled distance $x$ as

$$A(x) = \log_{10}(I(0)/I(x)) \tag{4}$$

The Bouguer–Beer–Lambert law relates the absorbance via

$$A = \varepsilon x c \tag{5}$$

to the attenuation coefficient $\varepsilon$, the optical path length $x$, and the concentration $c$ [74–77]. The absorbance of a medium can be calculated by summing up the absorbance contributions of its individual components. In principle, the proportionality of the absorbance and the concentration is only valid for low concentrations. Since the main absorbance of the scintillator components is at a short wavelength below 400 nm, the extinction at the wavelength region of interest is dominated by impurities. The concentrations of these impurities are very small. Therefore, we can assume the linearity between concentration and absorbance in the wavelength region of interest and add up the individual contributions of the single components [57].

Since the absorbance is inversely proportional to the attenuation length, the overall attenuation length $\Lambda$ can be obtained from $i$ individual contributions $\Lambda_i$ via

$$\frac{1}{\Lambda} = \sum_{i=1}^{n} \frac{1}{\Lambda_i} \tag{6}$$

where $n$ contributions of absorption or scattering length can be included. The formula can also be used if the contributions are determined separately for the different components in the mixture from individual measurements [57]. The attenuation length shows a dependency on wavelength.

The transparency of liquid scintillators is very sensitive to the exposure of scintillator components to air or high temperatures. Oxidized molecules can absorb light and lead to yellowing. Contact with materials such as stainless steel can catalyze the reactions. To minimize such yellowing, coated vessels of additions of antioxidants can be used [39].

Apart from the loss of light by absorption, scattering plays an important role. While scattered light is not lost, scattering can degrade the performance of vertex reconstruction. It can also indirectly cause absorption because a scattered photon can travel a longer path until it reaches a photosensor, i.e., it has a higher chance of being absorbed. In the absence of impurities, Rayleigh scattering dominates over Raman or Mie scattering. Rayleigh scattering involves scattering off bound electrons in the scintillator. It is strongly dependent on the photon wavelength $\lambda$, with additional dependence on the molecular polarizability $\alpha$:

$$I = I_0 \frac{8\pi^4 \alpha^2}{\lambda^4 R^2} \left(1 + \cos^2 \theta\right) \tag{7}$$

where $R$ is the distance to the particle, and $I$ and $I_0$ are the resulting intensities of the scattered light in direction $\theta$ and the initial intensity, respectively [78]. As can be seen from Equation (7), light is scattered symmetrically. The Rayleigh scattering length for typical solvents is a few tens of meters at around 430 nm [79].

*2.8. Purification*

The presence of quenching molecules or radioactive impurities can impact the performance of a scintillator by lowering their light yield or transparency and impose a significant background. Their removal can be achieved by a number of techniques, which can also be combined [39].

Dissolved quenching gases (e.g., oxygen) or radioactive noble gases can be purged from the scintillator using nitrogen [80–83]. It is also possible to reduce the solubility of gases by increasing the temperature or reducing the partial pressure of the vapor [84]. Alternatively, the use of ultra-sonic excitations allows the scintillator to be degassed.

Through distillation, less volatile components of the scintillator can be removed by exploiting their boiling points [80–82,85]. The process can be controlled via temperature and pressure. It is more effective to use distillation on pure solvents, as this avoids separation of already dissolved fluors. Typically, the last 10% of solvent containing all collected impurities has to be discarded.

Using the technique of water extraction, impurities are transferred to an immiscible aqueous phase in contact with the scintillator due to relative solubilities [80,81,86]. This technique allows a removal of radioactive impurities mostly coming from fluors. It is a continuous purification technique without the loss of the scintillator.

Filtration is another continuous purification technique [80]. Here, particles above a certain size, typically in the order of sub-micrometer, are removed. This technique is suitable to extract radioactive contaminants or other impurities.

Another approach is column purification, where the scintillator is sent through an absorber material such as aluminum oxide or silica gel [56,58,81,87]. The process improves the attenuation length of the scintillator. It is possible to use several layers of absorber with varying pH values to attack various impurities.

*2.9. Cherenkov Light*

Scintillation light is emitted isotropically from the point of production. There is also a directed component, Cherenkov light. Its contribution to the overall light output is energy- and particle-dependent and roughly a few percent. Cherenkov light is emitted through the Vavilov–Cherenkov effect [88], when the speed of a charged particle exceeds the speed of light in a dielectric medium. The frequency spectrum of Cherenkov radiation by a particle is given by the Frank–Tamm formula:

$$\frac{\partial^2 E}{\partial x\, \partial \omega} = \frac{q^2}{4\pi} \mu(\omega) \omega \left(1 - \frac{c^2}{v^2 n^2(\omega)}\right) \text{ for } \beta = \frac{v}{c} > \frac{1}{n(\omega)} \tag{8}$$

where *E* is the amount of energy emitted, per unit length *x* and per frequency $\omega$, $\mu(\omega)$ is permeability, $n(\omega)$ is the index of refraction of the material that the charged particle moves through, *q* is the electrical charge of the particle, *v* is speed, and the speed of light in vacuum is *c* [89]. Typical values for *n* in a liquid scintillator are around 1.4, but highly frequency-dependent. Higher frequencies are more intense in Cherenkov radiation such that most Cherenkov radiation is in the ultraviolet range. Cherenkov light is emitted in a cone, whose angle depends on the particle kinetic energy and the index of refraction. It becomes narrower with the higher speed of the particle. Part of the Cherenkov light, especially in the high-frequency region, is absorbed by the scintillator solvent or fluors. It is then reemitted isotropically at the emission spectrum of the fluors.

## 3. Metal-Loaded Scintillators

In many applications, it is beneficial to load a scintillator with metals. For example, boron or gadolinium allow for the efficient capture of neutrons in a scintillator [39]. Metals such as tellurium are suitable target materials in the search for neutrinoless double beta decay [90].

The challenging aspect of metal loading is the creation of a stable solution containing the inorganic metal. Chemical complexes that dissolve in a non-polar organic scintillator and do not diminish its optical properties are rare. Typically, these complexes are large compared to metal, such that the effective loading of the metal is much lower than the loading with respect to the complex. In addition, the complexes are quenchers that reduce the light yield of a scintillator. This can be partially counteracted by the use of high concentrations of fluors that compete with the absorption of the complex [43].

One possibility for metal loading is the use of a solvent with high solubility for polar compounds. In the CHOOZ experiment, gadolinium salt was dissolved in the alcohol hexanol in a first step [72]. In a second step, the loaded hexanol was diluted in other organic solvents. However, the nitrate ions of the salt deteriorated the attenuation length in the experiment.

Another approach is the use a surface-active agent (surfactant) with hydrophilic and hydrophobic chemical groups. Using this approach, the metal can be loaded into the aqueous phase, which is then brought into the scintillator phase. This approach was used by PROSPECT in the context of lithium-loading [60].

Other approaches involve organo-metallic complexes. A common choice is carboxylates [54,64]. Already in the first neutrino experiments, cadmium octoate (2-ethylhexanoic acid) was used in the 1950s [49]. Palo Verde used gadolinium 2-ethylhexanoate [53]. More recently, Daya Bay and RENO used 3,5,5-trimethylhexanoic acid (TMHA) [62,63].

An alternative to carboxylic acid systems is beta-diketones. They are expected to be more stable than carboxylic acid systems. Moreover, their stability above 200 °C and their high vapor pressures ease the purification processes. Beta-diketones have been researched for indium and zirconium loading and were applied in the context of gadolinium-loading in Double Chooz, Nucifer, and STEREO [18,56,57,91,92].

A more exhaustive review of metal loading techniques in liquid scintillators can be found at reference [39].

## 4. Blended Scintillators

In some situations, it is beneficial to use a mixture of solvents instead of a single solvent in a liquid scintillator [18,39,51–53,55,56,72]. This can be driven by transparency, matching between densities or light yields of neighboring detector sub-volumes, or by material compatibility arguments, which exclude the use of chemically aggressive solvents. In such situations, one can dilute the aromatic solvent in mineral oil or n-alkanes. This approach can also help to meet safety requirements with respect to the flash point of the scintillator.

The addition of non-scintillating solvents can also allow the scintillator light yield to be matched across optically connected volumes containing different scintillators [56]. Some additions of solvents improve the pulse shape of the scintillator. In metal-loaded scintillators, the admixture of solvents with a similar absorption spectrum as the metal complex allows one to mitigate the quenching effects of the complex to a higher degree than is possible with high concentrations of a competing fluor alone [57].

## 5. Low-Temperature Scintillators

An attempt to achieve sub-percent energy resolution in a liquid scintillator detector is undertaken by the JUNO collaboration in their satellite experiment TAO [93]. TAO, the Taishan Antineutrino Observatory, will be a Gd-loaded tonne-scale liquid scintillator detector placed about 30 m away from one of the cores of the Taishan Nuclear Power Plant. The measurement will provide a reference spectrum for JUNO. To achieve superb performance, an optical coverage of close to 95% using novel silicon photomultipliers (SiPMs), with a photon detection efficiency above 50% is intended. Approximately 4000 SiPMs are going to cover an area of about 10 m$^2$. The resulting photoelectron yield is expected to be about 4500 per MeV, i.e., an order higher than for typical large-scale liquid scintillator detectors.

To efficiently reduce the dark count of the SiPMs, the detector will be cooled down to $-50\,^\circ$C. LAB, which will be used in TAO, is mostly a mixture of molecules with 9 to 14 carbon atoms in the linear chain. It has a freezing point below $-60\,^\circ$C such that it remains a liquid. However, at a detector temperature of $-50\,^\circ$C, several other complications with respect to the scintillator become imminent. The water content in LAB may precipitate and result in transparency degradation. This can be counteracted by extensive drying through bubbling with dry nitrogen to remove water [94]. Furthermore, the solubility of fluors and wavelength shifters is largely reduced at low temperatures. Currently, an effective approach to this problem is the addition of dipropylenglycol-n-butylether (DPnB) in sub-percent quantities as a freezing inhibitor and antioxidant, which improves solubility [66]. However, since ethers tend to show peroxide formation over time, the introduction of DPnB could also affect the long-term performance through increased quenching. Further studies in this respect are ongoing.

## 6. Water-Based Liquid Scintillators

Monolithic optical detectors, either water–Cherenkov detectors or liquid scintillator detectors, are a well-established technique in neutrino physics. Using water-based liquid scintillators (WbLS) is an approach that exploits Cherenkov and scintillation signals simultaneously; i.e., water is loaded with 1% to 10% liquid scintillator [95–98]. The approach is similar to previously used highly diluted organic scintillators [70,71] or an approach using slow scintillators, which will be discussed in Section 7. However, since water is the main component in WbLS, the WbLS approach offers benefits such as low costs and strongly reduced fire and environmental hazards. The WbLS technology is foreseen in the Eos and ANNIE experiments [99,100] and could be deployed in planned experiments such as AIT-NEO, Theia, and beyond [20,101,102] or in the context of medical particle-beam therapy [103]. The technology is expected to allow improvements in many fields, such as high-energy, nuclear, geo-, and astrophysics such as neutrino mass ordering, CP violation in the leptonic sector, solar neutrinos, diffuse supernova neutrinos, neutrinos from supernova bursts, neutrinos from the Earth's crust, nucleon decay, and neutrinoless double beta decay with sensitivity towards normal neutrino mass ordering [104–108].

In a water-based liquid scintillator detector, it is possible to use the Cherenkov signals to provide directional and topological information while maintaining the good energy resolution of liquid scintillators. This separation between Cherenkov and scintillation light has been demonstrated in the CHESS setup [109,110]. Water-based liquid scintillators are expected to have good particle-identification capabilities (PID) following a discrimination strategy based on the particle-dependent Cherenkov/scintillation light ratio. This PID can improve the discrimination of alpha/beta particles and might allow some discrimination of beta/gamma particles. The PID performance of WbLS will be investigated by Eos [99]. It arises from two sources: the time profile of scintillation light emitted due to a recoiling proton may differ from electron-like events due to quenching effects, and the ratio of Cherenkov to scintillation light differs between heavier and lighter particles. Additionally, recent developments have demonstrated the capability to identify neutron/gamma particles through the pulse shape discrimination of the scintillation light [111]. Another additional benefit exists with respect to metal loading. In WbLS, loading can happen in the aqueous phase, which is easier to achieve than the direct loading of the organic scintillator (cf. Section 3).

The separation of Cherenkov and scintillation light signals can be achieved by two means. One idea is to separate fast-Cherenkov and slow-scintillation light time-wise [112]. This is in particular achieved via fast photon detectors, e.g., a Large Area Picosecond Photo-Detector (LAPPD™) [113–115]. Another idea is spectral separation between long-wavelength photons, which are dominated by the Cherenkov light, and short-wavelength photons, which can be – dependent on the scintillator fraction in the detector medium – dominated by scintillation light. Here, photodetectors with strong wavelength-dependent efficiency or dichroicons, Winston-cone-style light concentrators built out of dichroic reflectors, can be used [116,117].

Organic solvents, as used in liquid scintillators, are immiscible with water. This is mainly caused by the differences in the polarities of the molecules. To produce water-based liquid scintillators, the hydrophobic (lipophilic) scintillator component has to be brought into a stable suspension with the hydrophilic (lipophobic) water phase. An ampliphilic surface-active agent (surfactant) consisting of molecules with lypophilic and hydrophilic groups can be used to emulsify the organic solvent into the water solvent by reducing the tension between the organic solvent and the water [118]. The degree of tension reduction depends on the concentration of the surfactant. Its concentration can in turn also affect the optical and stability properties of the medium. A typical surfactant molecule possesses hydrophilic groups on one end and hydrophilic groups at the opposite end of the molecule. They can therefore form a hydrophilic shell around a hydrophobic droplet of scintillator inside the water, a so-called micelle. Micelles of large size can cause substantial translucence. Likewise, a high concentration of micelles can give rise to opacity.

Early studies investigated the possibility of producing water-based liquid scintillators from linear-alkyl-benzene-sulfonate (LAS), a derivate of the well-known linear-alkyl-benzene (LAB) [95]. Its light yield was found to have a dependence on the scintillator concentration of $(127.9 \pm 17.0)$ photons/MeV/%LS and an intercept value of $(108.3 \pm 51.0)$ photons/MeV, indicating a non-linear behavior at low concentrations [119]. For scintillator fractions between 1% and 10% in water, a clear dominance of Cherenkov light over scintillation light in the rising part of the light pulse could be seen in a fit the data (cf. Figure 2). From about 5% loading upwards, the fraction of scintillation light starts to dominate the peak of the pulse by more than an order of magnitude. A measurement of the relative proton light yield of a 5% WbLS showed it to be approximately 3.8% lower than that of a pure LAB+PPO reference [120].

An alternative approach to WbLS uses 13% Triton™X-100 as surfactant combined with 86% water and 1% LAB including 100 g/L PPO, as well as 10 mg/L vitamin C for pH control [98]. Here, a light yield of about $(198 \pm 5)$ photons/MeV is reported. The distribution of the sizes of micelles peaks at 2.8 nm. An important step in the production is the filtering of the WbLS, typically at the order of 10 nm to 100 nm.

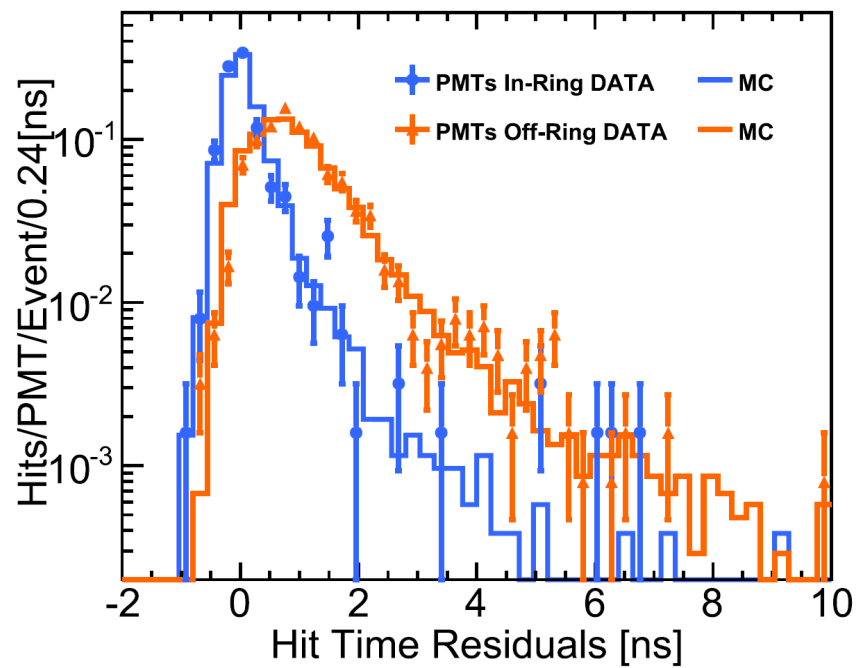

**Figure 2.** Data to Monte Carlo (MC) comparison of PMT hit-time residual distributions for PMTs inside and outside of a Cherenkov ring measured in 5% WbLS with the CHESS setup [109]. Reprinted from [119] under CC-BY license.

## 7. Slow Scintillators

The slow scintillator approach follows the same idea as the WbLS approach discussed in Section 6. By separating Cherenkov and scintillation light, advantages of both detection techniques can be exploited. In the slow scintillator approach, this separation is achieved via time-wise separation of the fast Cherenkov component from the slower scintillation component [112,121]. Slow scintillators are expected to show a good PID for proton/electron separation and some separation between electrons/gammas, as well as a vertex position resolution below 10 cm at energies above a few MeV [122,123]. The Cherenkov/scintillation separation can be achieved in different ways.

An early approach used pure LAB without fluors [124,125]. Due to the missing fluors, the intrinsic time constant of the LAB is already sufficiently slow to allow an effective separation. However, the light yield in pure LAB is far inferior to fluor-loaded scintillators, which makes this first approach vastly unsuited for neutrino applications.

An alternative is given by introducing only a small concentration of primary fluors to the LAB solvent [126]. The non-radiative energy transfer to the fluor strongly depends on its concentration, as discussed in Section 2.1, and is therefore inhibited, which renders the Cherenkov signal rather prominent. A reasonable balance between the Cherenkov and scintillation light separation power, on the one hand, and the light yield, on the other hand, was found for LAB using 0.07 g/L of PPO and 13 mg/L of bis-MSB. It is not possible to maximize both properties at the same time, because light yield and decay time constants were found to follow an inverse relationship. For the used fluor concentrations, a light yield of 4000 photons/MeV was found, and an attenuation length of 20 m seems achievable after purification.

Yet another idea is given by an approach to use slow fluors, i.e., a departure from the classical fluors PPO and bis-MSB in combination with LAB, in favor of intrinsically slow fluors [127]. For this approach, four fluors were investigated in LAB, two primary fluors (acenaphthene and pyrene) and two secondary fluors (9, 10-diphenylanthracene (DPA), and 1, 6-diphenyl-1, 3, 5-hexatriene (DPH)), which were combined with PPO. This approach was especially devised to overcome the potential difficulty of time-wise separation in the water-based approach, namely that wavelength dispersion in large detectors will broaden the prompt Cherenkov signal, impeding the signal separation. It is also suited to overcome the loss in light yield and therefore energy resolution in the approach using low concentrations of classical fluors. It was found that the selected fluors yield either comparable or superior light yield and separation power when compared to the approach of low concentrations.

## 8. Opaque Scintillators

The main idea behind opaque scintillators is the confinement of scintillation light around its production point by reducing the scattering length of photons in a material to below the cm-level, while keeping the absorption length high enough to ensure a good light output. This approach would preserve individual energy depositions, e.g., Compton-scatter vertices of a gamma-particle, and thereby allows to perform particle identification using the topology of energy depositions of events (cf. Figure 3). Opaque detectors could therefore have the ability to separate positrons from electrons and gammas on an event-by-event basis, reducing, e.g., the necessity of overburden [128]. Moreover, due to the confinement of scintillation light, the technology could allow tracking of particles above several MeV and their identification based on topology without or in addition to conventional track bending by a magnetic field. In order to collect the confined light, an opaque scintillation detector has to be instrumented with optical fibres that collect the scintillation light at the interaction point and transport it to SiPMs. The readout time resolution provided by the SiPM could give high precision for reconstruction in the direction along the fibre such that an unidirectional instrumentation with fibres appears feasible. The position reconstruction along the fibre can further be improved by implementing two alternating population of fibres, which are slightly tilted against each other [129].

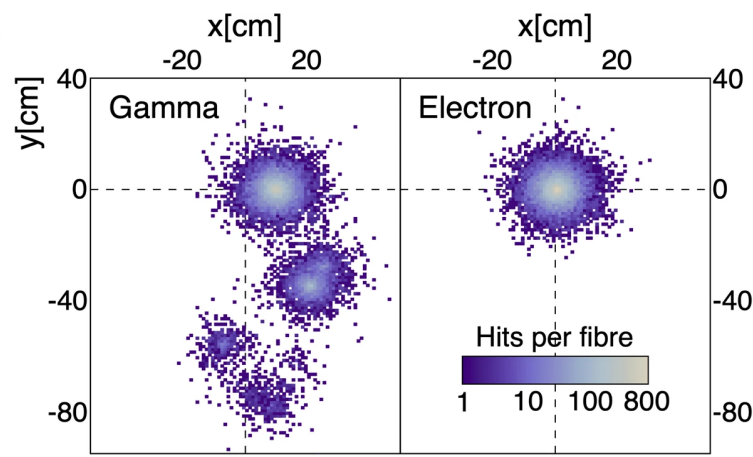

**Figure 3.** Simulation of a gamma (**left**) and electron (**right**) event with 2 MeV kinetic energy. Fibers are arranged along the *z*-direction in a lattice of 1 cm pitch. While the electron shows a single energy deposition, the gamma deposits energy at several vertices through Compton-scattering. A positron (not shown) combines both patterns. Reprinted from [128] under CC-BY license.

As the first opaque detector medium, the wax-based NoWaSH was examined in more detail in 2019 [130]. It combines the well-known mixture of LAB and PPO with up to 20% paraffin wax. At a temperature of 60 °C, the components can be mixed well before they solidify at below 20 °C to form a colorless, opaque solid. This method also enables complicated detector geometries to be filled in the liquid phase and, at the same time, offers additional protection against loss of NoWaSH through leakage after solidification, which makes it suitable for applications in nuclear facilities or underground laboratories. Many properties of NoWaSH are comparable to the main ingredient LAB. The refractive index and also the kinematic viscosity are similar to those of LAB, so that pipe and pump systems for LAB can easily be used after installation of additional heating. The paraffin wax used in NoWaSH exhibits good radiopurity. Activities of common contaminants were constrained to $\mathcal{O}(\text{mBq/kg})$ or below. Thus, NoWaSH could be used in low-background experiments, if other detector components, especially optical fibers running through an opaque detector, can achieve acceptable radiopurities. Absorption and emission spectra of NoWaSH are comparable to those of LAB, and the light output of NoWaSH is above 80% compared to LAB despite an inactive wax component of up to 20%. A major design difference is the scattering length. Values in the millimeter range could be achieved. NoWaSH thus opens up the possibility of future experiments with highly loaded scintillators of $\mathcal{O}(10\%)$ or more because of reduced requirements for transparency as compared to classic scintillators. Since NoWaSH solidifies when becoming opaque, new direct metal loading techniques without the use of a chemical complex (cf. Section 3) can be applied. In addition to NoWaSH, other ideas of possible opaque scintillators via suspensions or colloids have been proposed [131].

A proof-of-concept of the opaque detector technology was performed in the Micro-LiquidO detector, with an active volume of 0.2 L [128]. Here, electrons with an energy of 1 MeV were injected in the detector, filled with NoWaSH and instrumented with optical fibers. Compared to the transparent scintillator, more light was collected by the fibers near to the entry point than on rear fibers. This effect increases with increasing opacity, but does not lead to a loss of light overall. The successor detector Mini-LiquidO with a volume of 7.5 L presented the first results on the light confinement and full light ball formation in 2022 [129]. Mini-LiquidO was also able to achieve the time-wise separation of Cherenkov and scintillation light with its fast SiPMs using a slow scintillator. In 2022, a pathfinder phase for the proposed kilotonne-scale detector SuperChooz started. In 2023, a project for reactor neutrino measurements with the tonne-scale detector AMOTech/CLOUD will commence [132]. Opaque scintillator technology was also proposed for the detection of solar neutrinos using indium, geoneutrinos, accelerator neutrinos, neutrino oscillations, and neutrinoless double beta-decay [128,132–134]. However, most of these ideas have not been worked out in detail yet.

## 9. Siloxane-Based Scintillators

In several cases, liquid scintillator detectors are located at sites with enhanced safety requirements. These include underground laboratories or setups in the vicinity or inside nuclear-reactor buildings. Thus, there is strong interest in the reduction of hazards related to environmental pollution via leakage, toxicity, and flammability. In particular, properties such as the flash point and vapor pressure, in addition to the toxicity, are highly relevant. In that respect, there have been a number of new materials suggested and introduced in the field of liquid scintillators. A particularly promising suggestion from 2015, which recently received new attention, is polysiloxane-based liquid scintillators [135]. They were originally proposed in solid forms in 2010 [136]. Polysiloxane-based liquid scintillators exhibit outstanding properties in terms of safety. They have a low vapor pressure resulting in odourlessness and absence of harmful vapors. Their flash point is typically well above 200 °C. The viscosity of the polysiloxanes can be as high as 10,000 mm$^2$/s at room temperature. While this reduces the risk of spills and leakage, liquid handling procedures, such as pumping and filtering, becomes challenging. An advantage over classical scintillator liquids with C–C bonds is the much stronger Si–O bonds in the main chain of the polysiloxane molecules, which render them chemically inert. This positively affects their stability and good material compatibility. The formation of free radicals, which cause yellowing in organic materials, is less likely.

The most promising compound within the group of polysiloxanes is tetraphenyl-tetramethyl trisiloxane (TPTMTS). It has a high molar fraction of phenyl groups and a favorable absorption and emission characteristics, resulting in a high light yield. The fluor concentrations needed for polysiloxanes are higher compared to classical liquid scintillators. This might be explained by higher viscosities because the energy transfer from processes in which diffusion plays a role is hindered. TPTMTS is commonly used as a diffusion pump oil, given its rather low viscosity among liquid polysiloxanes, which also makes it easier to handle in the context of liquid scintillator detectors. In the literature, light yields of TPTMTS are reported to be comparable to LAB or the commercial xylene-based EJ309, which roughly corresponds to 80% anthracene [135,137]. These values could also depend on varying purity of the measured specimens. In terms of pulse shape discrimination (PSD), properties are similar to LAB and the attenuation length was measured to be better than 5 m [137].

## 10. Quantum Dots

Since their inception in the 1980s, quantum dots have become a well-known technology and have received new attention in the field of neutrino scintillation detectors during the last decade. Quantum dots (QD) are nanometer-scale objects, typically constructed from a semiconductor material [138]. Because charge carriers inside the semiconductor, such as electrons and holes, are spatially confined in a quantum dot, their energy states take on discrete values. Thus, quantum dots behave similarly to single atoms. The decisive difference from atoms is that their electronic and optical properties can be customized, because smaller dots have a larger band gap and vice versa. In the context of scintillators, the adjustability of light absorption and emission spectra makes them unique candidates for fluors. Due to the quantum nature of the emission, the spectra are narrow resonances around the characteristic wavelength determined by the sizes of the dots. Likewise, tuning the absorption spectrum could offer a way to separate Cherenkov and scintillation spectra in the context of water-based liquid scintillators (cf. Section 6). Quantum dots might also offer a way to load scintillators with isotopes, e.g., for neutrinoless double beta decay searches. They are produced with an organic shell of ligands, which enables to suspend them in organic solvents.

The first studies used toluene with 5 g/L PPO as scintillator. Toluene-suspended quantum dots were added to the toluene-based scintillator as a secondary wavelength shifter at a concentration of 1.25 g/L [139]. As a material, CdS quantum dots were selected, because their emission wavelengths fall into the 360–460 nm matching typical PMTs' sensitivities. The light yield measurements showed a reduced light output of the quantum-dot-doped scintillator. The timing distributions for the quantum-dot-doped scintillators are similar to the standard toluene scintillator. In this study, it was found that the dark rate is higher for the quantum-dot-doped scintillator. It is speculated that the increased rate may be due to the process of quantum dot "blinking". Blinking denotes a photoluminescence intermittency in the quantum dot emission, arising from the escape of either one or both of the photoexcited carriers to the surface of the quantum dot [140]. In the first case, the residual charge of the quantum dot quenches photoluminescence via a non-radiative Auger recombination. If both carriers escape, the exciton is intercepted before thermalization and does not contribute to the photoluminescence. There are attempts to suppress blinking via non-radiative Auger recombination by weakening the confinement potential between core and the shell of the quantum dot. Another issue could arise because the organic molecules that allow the quantum dots to be suspended in organic solvents or water can cause clumping of dots and thereby a degradation of their performance. A degradation of transparency after several weeks, consistent with the clumping effect, was found but can be counteracted through filtering, such that attenuation lengths of several meters are possible [141]. A later study investigated quantum dots with a new type of structure, a perovskite [142]. It exhibits somewhat higher luminescence and ease of synthesis.

Before quantum dots can be used in scintillator detectors, higher light yields and improved stability must be achieved. There are significant efforts to improve quantum dots at industrial scales, which could allow quantum dots to be used in future liquid scintillator applications. There have also been studies to include quantum dots in waveguides [143], with possible applications to scintillation detectors.

## 11. Floating Liquid Scintillators

A novel idea with respect to the design of scintillation detectors was presented in 2022. The stratified liquid plane scintillator (SLIPS) detector [144] allows one to avoid the use of barriers that separate the scintillation region from the non-scintillating buffer region. This buffer region then accommodates photomultiplier tubes to avoid background signals from the intrinsic radioactivity of the PMTs to enter the scintillator region. The separation is often achieved through acrylic or nylon barriers, whose own activity can impose a problem for low-energy events in the MeV-range. In addition, the construction of these barriers can be challenging and expensive for larger detector volumes. As shown in an initial simulation study, by an adequate choice of liquids, physical barriers can be avoided. SLIPS uses layers of lipophobic non-scintillating liquids to separate photomultiplier tubes from the lipophilic scintillation region. If lipophobic liquids of different densities are used, a lipohobic layer above and below the scintillator can be achieved. Alternatively, an asymmetric design with one buffer layer at the bottom of the detector and reflecting walls on all other sides is proposed. As a buffer liquid, ethylene glycol is proposed, because its refractive index matches well with the scintillator liquid in the wavelength range of emission. The lateral sides of a SLIPS detector cannot be instrumented due to the missing buffer region in any design. The height of a SLIPS detector is limited by the requirement to be far shorter than the attenuation length of the liquids. As shown in the simulation study, this design performs reasonably well compared to classical detectors while using far fewer PMTs. The SLIPS design is inspired by the partial fill phase of the SNO+ detector. SLIPS might not be suited as a concurring design alternative for classical large-scale detectors, since excavation costs for underground caverns might be comparable. However, when adopting a long-box-shape design, it allows the opportunistic use of abandoned tunnels at the price of a more complicated vertex reconstruction. At the present time, prototype studies have only been conducted to a small extent. Here, the long term stability with respect to fluor transfer between liquids will be especially relevant.

## 12. Summary

In recent years, significant progress has been made in the context of organic liquid scintillators. Stable metal-loaded scintillators allow the use of liquid scintillators in a variety of applications in neutrino physics. The development of slow and water-based liquid scintillators offers to combine the benefits from previously separate detection strategies of Cherenkov and scintillation detectors. Opaque scintillator technology is expected to allow unprecedented particle identification and high metal loading. Advancements in stable cooled scintillators will enable superb energy resolutions in liquid scintillator detectors. Additional ideas for safe scintillators could allow measurements in sensitive locations very close to nuclear reactors. Customizable quantum dots could one day be used as tunable secondary wavelength shifters or for metal loading.

Several approaches discussed in this review are applicable to neutrino physics goals such as neutrinoless double beta decay, searches for neutrinos of solar, supernova, or astrophysical origin, as well as reactor and geological neutrinos. The active exploration of all approaches will be decisive to ensure further advances in neutrino physics.

**Funding:** This work has been supported by the Alexander von Humboldt Foundation funded by the German Federal Ministry of Education and Research (BMBF) as well as the Cluster of Excellence "Precision Physics, Fundamental Interactions, and Structure of Matter" (PRISMA$^+$ EXC 2118/1) funded by the German Research Foundation (DFG) within the German Excellence Strategy (Project ID 39083149).

**Institutional Review Board Statement:** Not applicable.

**Informed Consent Statement:** Not applicable.

**Data Availability Statement:** Not applicable.

**Acknowledgments:** The author thanks Javier Caravaca Rodriguez for the invitation to write this review article.

**Conflicts of Interest:** The author declares no conflicts of interest. The funders had no role in the writing of the manuscript or in the decision to publish it.